# Calming traffic in unsignalized bidirectional streets by introducing one-lane bottlenecks

Victor L. Knoop and Carlos F. Daganzo

## Abstract

A common strategy to calm traffic in an unsignalized, bidirectional street consists in narrowing the street to a single lane at one or more locations. These locations become bottlenecks. Two policies are commonly used there: first-in-first-out (FIFO), and directional priority (DP), which means one of the directions has priority. This paper examines the performance of these two policies on streets with one or more bottlenecks. Two measures of performance are considered: the expected delay (evaluated for typical situations with low demand) and the system capacity. The latter is defined as the pair of flows (d1, d2) that depart the street in the two opposite directions when the system is oversaturated with queues. For a street with a single bottleneck, the capacity under both FIFO and DP is not unique: In both cases, the values of (d1, d2) form a continuous decreasing curve. As expected, we find that arrival flow pairs (a1, a2) below this curve can get through the system without generating queues. Those above cannot. For a single bottleneck it is shown that FIFO yields lower expected delay than DP under low demand. For high demand, DP's capacity exceeds FIFO's when the flows are balanced, but are lower otherwise. For multiple bottlenecks, the expected total delay on the street under low demand is that of a single bottleneck multiplied by the number of bottlenecks. The multi-bottleneck capacity under FIFO turns out to be the same as for a single bottleneck. For DP, traffic near capacity is complex. Capacity points turn out to be numerous and disjointed. Formulas are derived for their values. Higher capacities are typically obtained for longer streets. Curiously, these capacity points are often above the capacity curve of a street with a single FIFO or DP bottleneck, indicating that an extra bottleneck can increase capacity.

## 1. Introduction

In The Netherlands, bottlenecks are often introduced on local two-lane bidirectional streets to calm traffic. Typically, these bottlenecks consist of a barrier that blocks the street except for one lane (see figure 1a, CROW 2026). The bottlenecks are unsignalized, so conflicts are resolved by rules-of the-road policies. Two are commonly used: FIFO and directional priority (DP). Under DP, one of the directions is assigned priority by a posted sign (see Figure 1.)

The bottleneck strategy seems to work well and calm traffic, but no systematic studies of its side effects have been published to date. This paper attempts to fill this void by investigating the strategy’s delays and capacity under both policy types.

To compare these two policies and understand the basic issues, we analyze a two-lane bi-directional street, spatially symmetric in the two directions, with several bottlenecks; and with all the traffic entering and leaving the street at its two ends; i.e., with no flow entering or exiting the system mid-street.

The paper will examine first, in Section 2 below, a light traffic scenario where queuing is rare, because this is when the traffic calming policies under study make the most sense. Formulas for delay will be derived. A focus on delay is justified because delay correlates closely with the costs a policy imposes on both drivers and non-drivers. The former ones suffer the delays directly, and the latter suffer the externalities of delay such as emissions and noise.

The paper will then examine, in Section 3, how the two policies perform under heavy traffic. For this scenario, the focus will be on the system's capacity rather than delay because under heavy traffic, capacity is the main determinant of delay—and accurate formulas for delay are elusive. The section will also explain how rough estimates of delay can be obtained from the balance between demand and capacity.

# 2. Light Traffic: Delay

This section develops formulas for the expected delay that our two calming policies produce under light traffic. Section 2.1 considers a street with a single bottleneck and Section 2.2 a street with any number.

## 2.1. A single bottleneck

It is difficult to obtain exact formulas for delay because the multi-vehicle interactions between the two opposing traffic streams are complex. However, approximations can be derived if traffic is so light that platoons and queues do not arise. This is done below.

First, a note on terminology: Unless stated otherwise, the opposing directions are labeled "1" and "2". Furthermore, to simplify some of the expressions, the unit of time (abbreviated [tu]) is chosen to be the average headway observed at saturation for a single lane in either direction; i.e., the average time between successive discharges from a single-lane queue at a traffic signal, and the average headway observed inside platoons. Thus, the per-lane saturation flow in both directions is 1. An estimate of [tu] from field measurements was 2.3 secs (+/-0.15 as standard error of the estimator).

For the derivation of delays that follow, it is assumed that the two opposing arrival streams are in a steady state. Arrivals are random and vehicles are assumed to behave identically. DP is analyzed first and FIFO second.

### *2.1.1. DP*

To begin, let us define the minimum time lag "$l$" required for safety by a non-priority vehicle between its bottleneck crossing and the arrival of the next priority vehicle. Let us also define the minimum time separation "s" between the arrival of a priority vehicle and the ensuing crossing of a non-priority vehicle that does not have to yield. The following estimates were obtained from field measurements: $l \approx 2.3$ [tu] (+/- 0.2) and $s \approx 1.8$ [tu] (+/- 0.15).

Clearly then, each priority vehicle arriving in isolation, blocks non-priority (or secondary) vehicles for a time $l$ before its arrival and a time $s$ after. The total duration of the blockage will be denoted $b = l+s$. This parameter will be called the "blocking time". Figure 1b is a time-space diagram that displays the bottleneck by means of a horizontal dotted line, an isolated priority vehicle by means of a rising solid line, and the blockage induced by this vehicle by means of a horizontal solid segment at the bottleneck location. The figure also includes the down sloping trajectories of two secondary vehicles. The first arrives before the blockage and goes through without delay. The second arrives during the blockage and must wait until it is over.

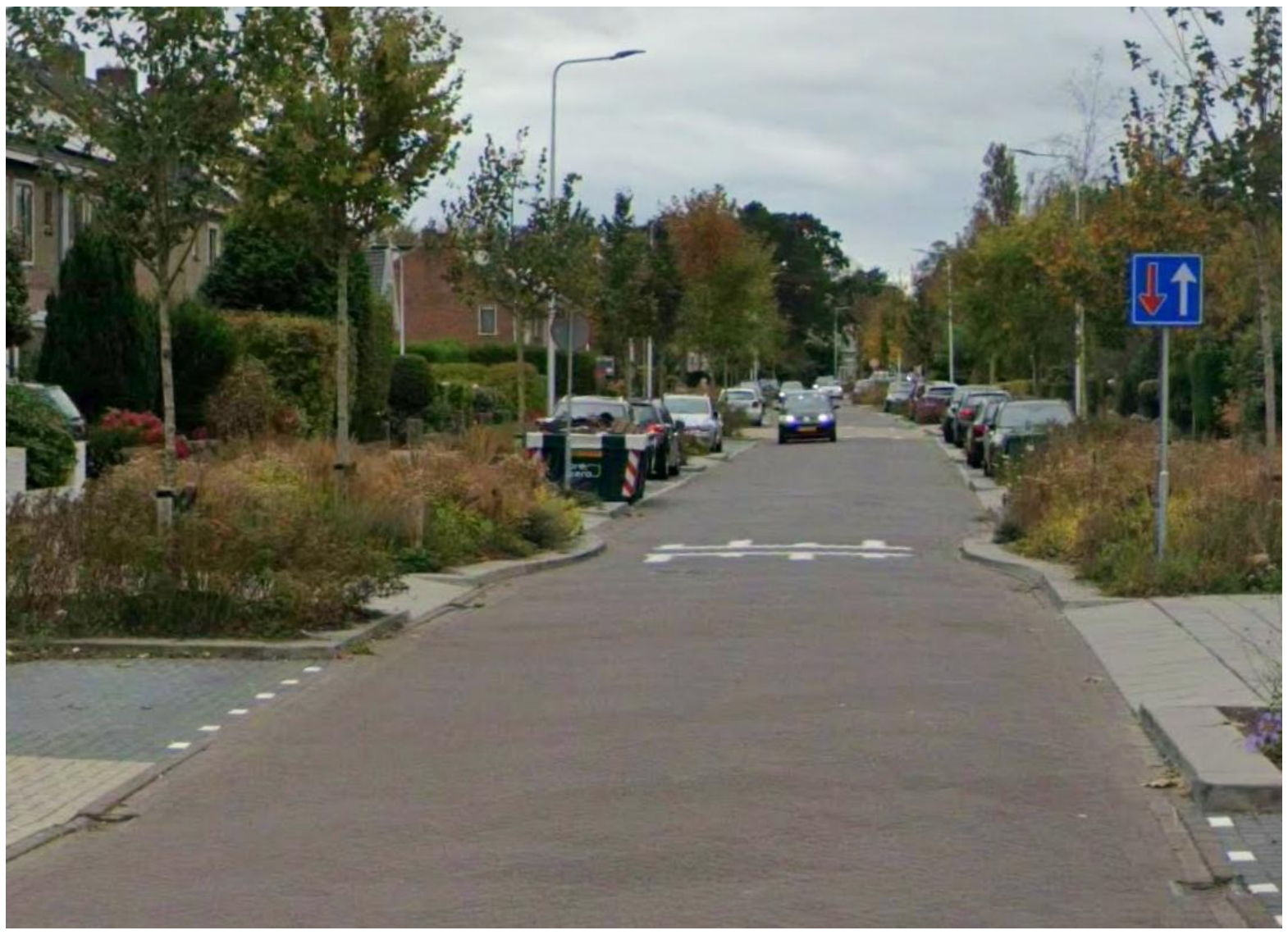

a) Physical appearance (Google maps)

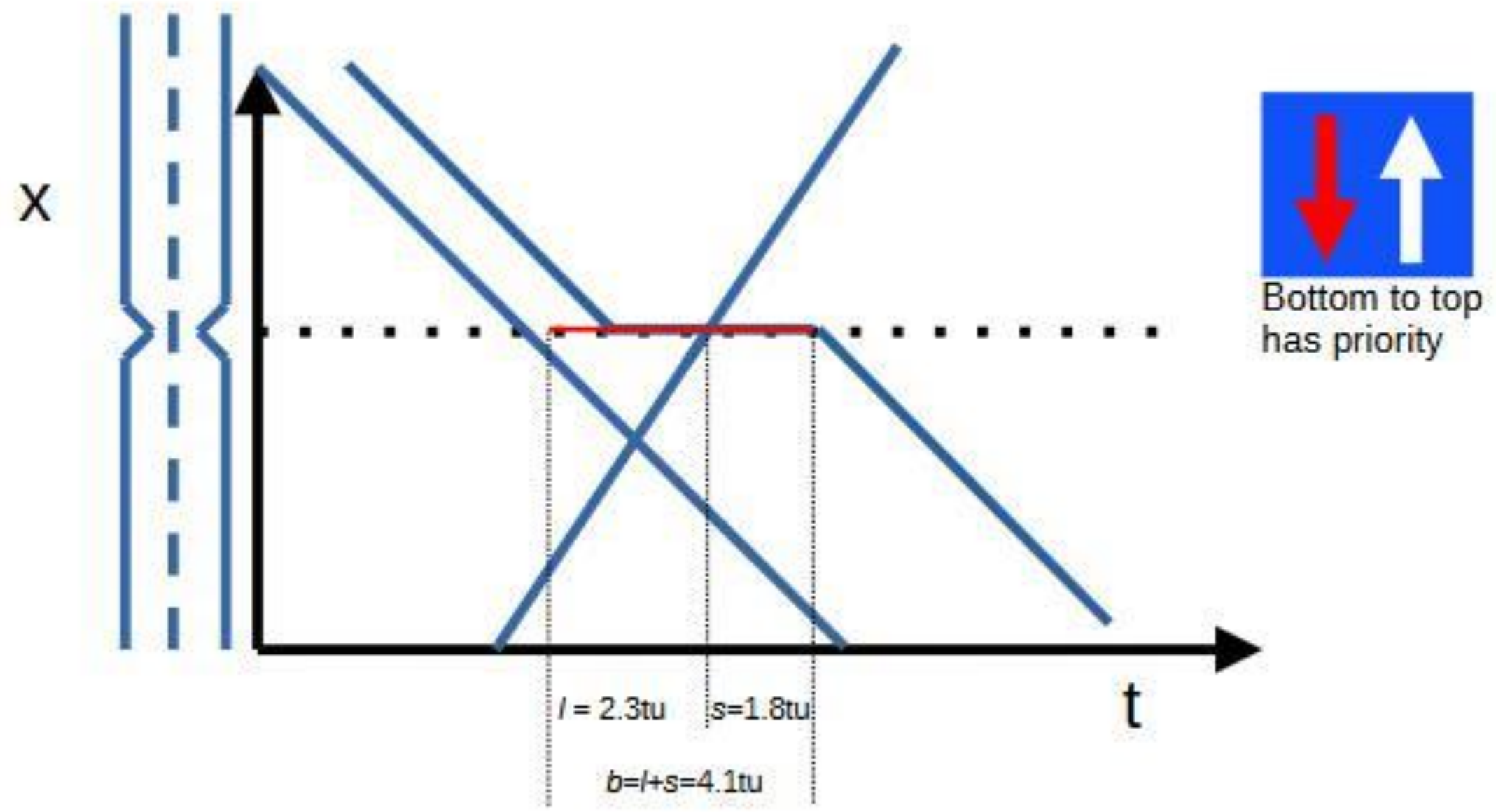

b. Time-space diagram showing how an isolated primary vehicle blocks secondary vehicles for a time *b= l+s*.

Figure 1: one-lane bottlenecks

We assume without loss of generality that direction 1 has the priority. Traffic is assumed to be so light in both directions that the expected number of arrivals in one blocking time in either direction is much smaller than one; i.e., $a_1 b$, $a_2 b$ << 1. This ensures that the blocking intervals of priority vehicles rarely overlap and that non-priority vehicles arrive to the bottleneck in isolation without having to queue. Since the blocking intervals of priority vehicles do not overlap, the expected delay of a non-priority (secondary) vehicle conditional on being blocked is b/2 [tu]. Furthermore, the whole stream of priority vehicles blocks the bottleneck for a fraction of time $a_1 b <<1$, which is also the probability of a randomly arriving secondary vehicle being blocked. The unconditional expected delay per secondary vehicle under DP, $D_{s, DP}$, is the product of the conditional expectation, b/2, and the probability of being blocked, $a_1 b$. Thus, we have:

$$D_{s, DP} \approx (a_1 b)(b/2)= a_1 b^2/2, \quad \text{where } (a_1 b, a_2 b) << 1, \quad \text{(delay per secondary vehicle).}$$

Now, let us express the expected total delay caused by the bottleneck in a time unit, $D_{T,DP}$ . Since only secondary vehicles experience delay, $D_{T,DP}$ is the product of the delay per secondary vehicle above and the arrival rate of secondary vehicles ; i.e.,

$$D_{T,DP} = a_2 D_{s,DP} \approx [a_2\, a_1](b^2/2), \quad \text{where } (a_1 b, a_2 b) << 1, \quad \text{(total delay per unit time).} \qquad (1)$$

Note how the above expression is symmetric to an interchange of $a_1$ and $a_2$. This indicates that the expected total delay under light traffic is independent of which direction is assigned priority.

*2.1.2. FIFO.*

Under FIFO there is no safety lag, but there still must still be a minimum time separation “s” between vehicles crossing the bottleneck in opposite directions.

Thus, every vehicle blocks the bottleneck to opposing traffic for s [tu] after it crosses the bottleneck. We continue to assume that traffic is light, in the sense that ($a_1$s, $a_2$ s) <<1, and that a blocked vehicle crosses the bottleneck when the blockage ends. Thus, a vehicle’s expected delay conditional on being blocked is s/2. For direction-1 vehicles, the probability of being blocked is the fraction of time their crossing is blocked by direction-2 vehicles, $a_2$s. Thus, the unconditional expected delay for direction-1 vehicles, $D_{1,v,FIFO}$, is:

$$\text{Direction 1: } D_{1,v,FIFO} \approx a_2 s\, s/2 = a_2 s^2/2 \quad \text{if } (a_1 s, a_2 s) <<1, \quad \text{(delay per vehicle).}$$

By symmetry, we can write for direction 2:

$$\text{Direction 2: } D_{2,v,FIFO} \approx a_1 s\, s/2 = a_1 s^2/2 \quad \text{if } (a_1 s, a_2 s) << 1, \quad \text{(delay per vehicle).}$$

Multiplying the above expressions by their respective flows per unit of time ($a_1$ and $a_2$) we obtain the expected total delay per unit time for each direction. Then, adding these, we obtain the expected total delay in both directions combined, $D_{T, FIFO}$:

$$\text{Both directions: } D_{T, FIFO} \approx [a_1 a_2]\,(s^2) \quad \text{if } (a_1 s, a_2 s) << 1, \quad \text{(total delay per unit time).} \qquad (2)$$

Equation (2) can be compared with (1) to see which policy results in less expected total delay. Note that the coefficient in square brackets involving the flows is the same for both equations. Thus, the determining factor is a comparison of the terms in parentheses. Using the values for s and b from field observations, we have: $s^2$=3.24 [$tu^2$]< $b^2/2$=7.2 [$tu^2$]. Hence FIFO yields about half the delay of DP. This is true regardless of the specific flows, or which direction is assigned priority.

## 2.2. Multiple bottlenecks

Consideration shows that if the arrival flows are as light as assumed in Section 2.1, we will not see any platoons forming regardless of the control policy. Then the logic leading to equations (1) and (2) also applies to each of the bottlenecks in our street. As result, these formulas and the qualitative conclusions they imply continue to apply to each of our bottlenecks. In other words, FIFO also yields less expected delay in the general case with multiple bottlenecks.

It turns out , as we show below, that FIFO does not perform as well under heavy traffic. Therefore, DP deserves serious consideration.

# 3. Heavy Traffic: Capacity

If bottlenecks are identical and do not interact then the flow that can get through one bottleneck can also get through the others; i.e. the capacity of a street with multiple identical bottlenecks should be the same as if it had only one bottleneck. Simulations confirm that this premise holds for FIFO, and also for DP if the priority direction is the same at all bottlenecks. In view of its importance, the single-bottleneck capacity is examined in Sec. 3.1 below. Section 3.2 then does the same for the only multi-bottleneck case (DP with changing priority directions) that is not equivalent to a single bottleneck.

## 3.1. Capacity: A single bottleneck

In what follows, the variables $x_1$ and $x_2$ are used to denote generically the flows in directions 1 and 2. Letters other than "x" will be used when we wish to characterize the flows more specifically; e.g., "($a_1$, $a_2$)" for arrivals, "($d_1$, $d_2$)" for departures, etc. We examine FIFO first and DP second.

### *3.1.1. Capacity: A single bottleneck under FIFO*

This subsection develops formulas for the bottleneck's departure rates ($d_1$, $d_2$) that arise when a persistent queue exists in either one or both sides of the bottleneck due to high, and possibly unbalanced demands ($a_1$, $a_2$).

For the analysis, without loss of generality, the persistent queue is assumed to arise in direction 1. Direction 2 is assumed to be undersaturated; i.e., with $d_2 = a_2$. In this direction the bottleneck may exhibit transient queues that alternate "busy" periods of queuing with "idle" periods without arrivals, departures or queues. It is convenient to introduce the parameter $f_2 \in [0, 1]$ to denote the fraction of time direction 2 is idle. As shown below, this allows us to express the capacities ($d_1$, $d_2$) in parametric form as a function of $f_2$. Note, the case with persistent queues on both sides of the bottleneck is also covered by these results, simply by setting $f_2 = 0$.

During periods of idleness, direction 2 is vacant without arrivals or departures (so $d_2 = a_2 = 0$) and the queue in direction 1 is unblocked (so it discharges at the unrestricted rate and $d_1 = 1$). During busy periods both directions are blocked and the two queues take turns discharging vehicles with the minimum separation under alternating directions, s. Thus, during busy periods, the discharge rates are $d_1 = d_2 = 1/(2s)$.

Now, given that direction 2 is idle for a fraction of time $f_2$ in the long run, and busy for a fraction $1-f_2$, our long-run capacity flows ($d_1$, $d_2$) must be:

$$d_1 = f_2 + (1- f_2)/(2s), \text{ where } f_2 \in [0, 1]; \quad (3a)$$

$$d_2 = (1- f_2)/(2s), \text{ where } f_2 \in [0, 1]. \quad (3b)$$

The loci of all points ($d_1$, $d_2$) obtained by varying $f_2$, is a curve in the ($x_1$, $x_2$) flow plane that defines all possible saturation flow-pairs when there is a persistent queue in direction 1—including also a persistent queue in direction 1 at the extreme point corresponding to $f_2 = 0$. The curve is a linear segment with extremes at points (1/(2s), 1/(2s)) and (1,0).[1]

By symmetry, the same formulas must apply with the subscripts interchanged if the persistent queue is in direction 2. Then we have:

$$d_2 = f_1 + (1- f_1)/(2s), \text{ where } f_1 \in [0, 1] \quad (4a)$$

$$d_1 = (1- f_1)/(2s), \text{ where } f_1 \in [0, 1] \quad (4b)$$

As expected by symmetry, (4) defines a linear segment with extremes at points (1/(2s), 1/(2s)) and (0, 1).

Equations (3) and (4) combined delineate the possible capacity points, regardless of where queues persist. For this reason, we call the resulting curve "the capacity boundary". Note, it is bilinear and convex-decreasing, with extremes at points (0, 1), (1,0) and a middle break-point at (1/2s, 1/2s). Figure 2 displays it for the experimentally observed value of s. As expected, the bilinear curve has mirror symmetry with respect to the line $x_1 = x_2$.

The smooth concave curve intersecting the bilinear curve and sharing its extreme points is the capacity boundary of the DP policy with priority given to direction 1. Its analytic expression is derived in Sec. 3.1.2 below. Note how the DP curve is not symmetric and slightly outperforms FIFO when $a_1$ and $a_2$ are balanced.

Arrival points ($a_1$, $a_2$) above either of the capacity boundaries are said to be oversaturated because they cannot be served without a growing queue. Conversely, arrival points beneath the boundary are undersaturated and cannot generate growing queues.

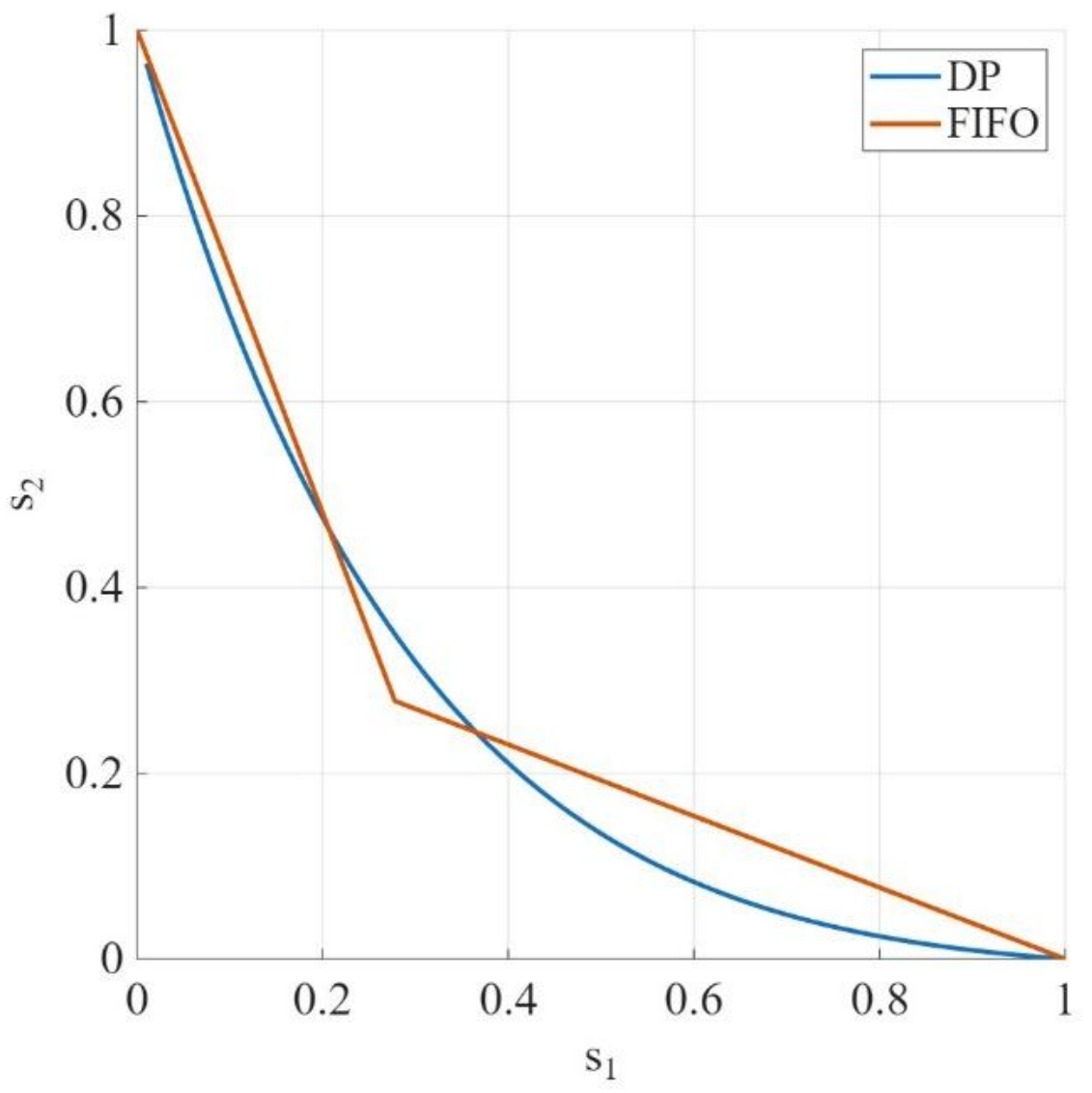

---

[1] The location of the segment's extremes should be intuitive because when $f_2 = 1$ direction 2 is permanently idle and direction 1 can discharge without obstruction, so then ($d_{1,}$, $d_2$) =(1, 0). And at the other extreme where $f_2 = 0$, there are queues on both sides, which must alternate discharges every s [tu], so ($d_{1,}$, $d_2$) = (1/(2s), 1/(2s)).

Figure2: Capacity boundaries of a single bottleneck under FIFO and DP (with priority to direction 1) for the observed b=4.1 and s=1.8. The boundaries intersect but have the same extreme points.

### *3.1.2. Capacity: A single bottleneck under DP*

The capacity boundary under DP is now derived. As in section 2.1.1, we assume that direction 1 has priority. Obviously then, this direction can have no queues and $d_1 = a_1$. We shall assume that direction 2 has a permanent queue and look for its discharge rate $d_2$ as a function of $a_1$, where $a_1 \in [0, 1]$. This function is denoted, $d_2 = c(a_1)$ to stress that the output is at capacity. The loci of points $(d_1, d_2)$ obtained for all $a_1 \in [0, 1]$ is the capacity boundary under DP.

It shall also be assumed that the arrival process of priority vehicles is random and stationary with average rate $a_1$. To avoid infeasibly short headways while preserving randomness, it will be further assumed that the arrival stream is the departure stream of an imaginary upstream M/D/1 queue with Poisson arrivals (at rate $a_1$) and deterministic service times of duration 1 [tu]. This setup implies that priority vehicles arrive in compact platoons of unit headways (H=1) and that the longer headway ahead of a platoon leader $H_L$ is a shifted exponential random variable, such that $H_L = 1 + E$, where E is negative exponential with mean $1/a_1$. Appendix A proves that under these conditions the capacity for direction 2, $d_2 = c(a_1)$, is given by the formula:

$$d_2 = c(a_1) = a_1 (1- a_1) \exp[-a_1 (b-1)] / [1-\exp(-a_1)]. \qquad (5)$$

Since $d_1 = a_1$, the capacity boundary under DP, shown in Figure 2, is the curve $(d_1, d_2) = (a_1, c(a_1))$, where $c(a_1)$ is given by (5).

### *3.1.3. How capacity affects delay under heavy traffic*

Under heavy traffic, the proximity and relative position of the point denoting the input flows $(a_1, a_2)$ to the capacity boundary critically influences delay. If the input point is feasible (i.e., beneath the boundary) all queues will be transient and the expected steady state delay on the two approaches combined should be finite. This expected delay, however, increases toward infinity as the input point approaches the boundary from below; i.e., as the bottleneck approaches saturation. When the system is oversaturated (i.e. the input point is above the capacity boundary) either one or two permanent queues will grow steadily, and the total system delay will also grow without bound.

Although formulas for delay cannot be given, it should be clear that during periods of high traffic, a priority strategy with high capacity should yield less delay than one with lesser capacity. Thus, knowledge of the capacity boundary for each priority policy can help authorities decide which policy to use during periods of high demand.

## 3.2. Capacity: General results for multiple bottlenecks

We propose that the capacity boundary of a multi-bottleneck street governed by either FIFO or DP with the same priority direction at all bottlenecks is the same as that of a single bottleneck. This is reasonable to expect because with these priority schemes, the bottleneck mechanisms are identical at all bottlenecks regardless of the traffic arrival patterns. The equivalence of multi- and single-bottleneck capacity boundaries is supported by simulations. To be complete, we now examine the remaining multi-bottleneck case: DP with different priority directions. Because this case turns out to be rather complex, we only consider in detail a street with two bottlenecks and where, as logic would dictate, priority is given to the (diverging) directions exiting the street to avoid vehicles being trapped in the section in between the bottlenecks.

## 3.3. Capacity: The special case of wo bottlenecks under DP with divergent priority directions

We will show that the two bottlenecks interact in a strange way that produces multiple capacity outcomes, and that quite often the outcome that arises sits above the capacity boundary of a single bottleneck.

From now on, the portion of the street between the bottlenecks is called "the block". To avoid internal queues that could cause gridlock, priority at our bottlenecks can be given in only two ways: (a) to the directions that point away from the middle of the block (diverging priority directions); or (b) to the same direction at both ends of the block (coincident priority directions). It was already pointed out in Sec. 3.2 that the capacity outcomes of case (b) should match those of a single bottleneck. Therefore, this section focuses on case (a). The following was found.

1. The capacity outcomes depend on block length. For blocks of any length, two capacity modes with very different capacities were identified. Mode A and Mode B. Simulations showed that these modes emerge spontaneously and unpredictably depending on how the system is loaded.

2. Mode A exhibits growing queues at both ends of the block and the traffic in between is bidirectional. Mode B exhibits transient queues that alternate discharges of the complete queue at both ends of the block, and the traffic in between is unidirectional, alternating in direction.

3. Under Mode A, the capacity is low for short blocks. For longer blocks, it is not unique: It can be high or low (see section 3.3.2)

4. Under Mode B, the system behaves as if the bottlenecks were governed by two actuated traffic signals that turn green when the queue at the other end has vanished and a gap of sufficient duration appears in that stream.

5. The capacity of Mode B is unique and predictable. For blocks of typical lengths, even if very short, the sum of the bidirectional capacity flows (i.e., the sum of the departing flows in both directions) approaches 1 as the duration of a simulation increases toward infinity.

Section 3.3.1 presents the simulation results supporting items 1-5 for blocks of different lengths. These simulations show the capacity outcomes and how they relate to the single bottleneck boundary under DP. Then, sections 3.3.2. and 3.3.3 analyze the two modes in detail. The results are consistent with the simulations. This consistency is valuable because it validates both the simulation and the analysis. The latter is also useful for the insights it produces.

### *3.3.1 Capacity of two bottlenecks under DP: Simulation results*

Figure 3 shows the capacity outcomes on the flow plane for the input parameters we observed in the field: $s$ = 1.8 [tu] and $l$ = 2.3 [tu]. The four panes correspond to blocks of different lengths: T=2.5, 3, 10 and 200 [tu]. Each simulation in the figure involved thousands of vehicles and is represented by a single light blue segment. Its top end marks the input arrival flows and the bottom end the output flows (capacity).

When the block is undersaturated, so the departure rates roughly match the arrivals, the segment is very short and displayed as a point.

For comparison purposes, the four panes also display the single-bottleneck capacity boundary under DP of Figure 2. The following qualitative observations are worth noting.

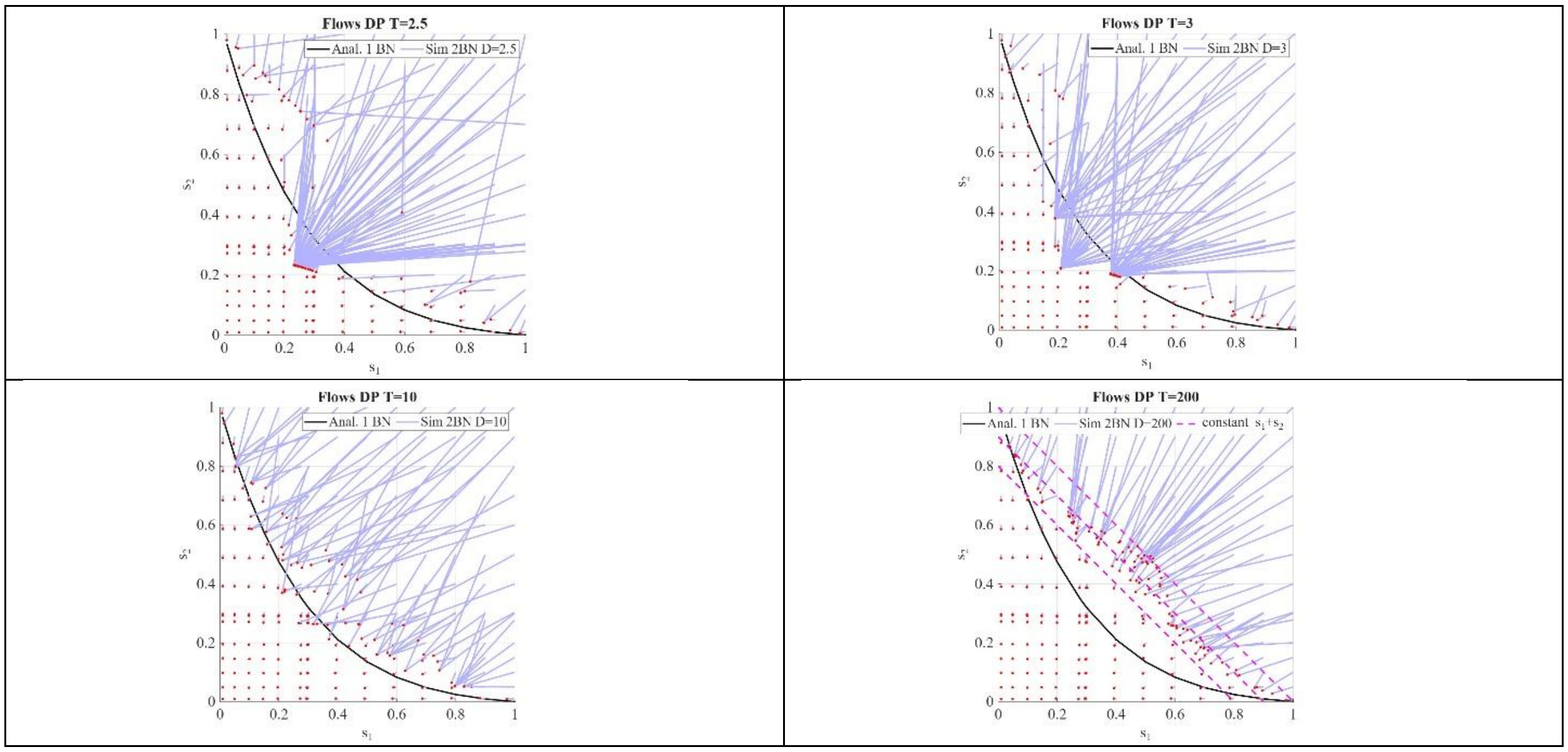

Figure 3: capacity outcomes for different block lengths

a. In all four panes, when arrivals are undersaturated for one DP bottleneck (i.e., the top of the segment is below the capacity boundary) then the arrivals are also undersaturated for two bottlenecks (i.e., the segment is a point). Although there are undersaturated outcomes above the capacity boundary, this is rare.

b. For short blocks (T=2.5), the capacity of oversaturated arrivals has an even chance of being higher or lower than with a single bottleneck. When it is lower the capacity points cluster together. This is Mode A (see item 3 above). The high-capacity points with bidirectional capacity close to 1 correspond to Mode B (see item 5 above); these points are found on a diagonal line (mostly in the left top of the figure).

c. Both modes also arise for medium blocks (T = 3), but in this case Mode A exhibits 3 clusters, which is consistent with item 3 above. All three clusters are below the DP capacity boundary.

d. For long blocks (T=10 and T = 200), Mode B is rare but the capacity flows of Mode A are non-unique and fairly high--consistently above the capacity boundary for DP. For unrealistically long blocks (T=200) the bidirectional capacity of Mode A is even higher, exceeding 0.8. For clarity of the values, we added dashed lines of iso-bidirectional capacity at 0.8, 0.9 and 1.

Sections 3.3.2 and 3.3.3 below examine Modes A and B in analytic detail.

### *3.3.2 Capacity of two bottlenecks under DP: Mode A*

We focus initially on the case for T=2.5 (short blocks) because it is easier to describe and understand. The left pane of Figure 4 shows a time-space diagram for the block with one bottleneck (named U for “up”) at the top and another (named D for “down) at the bottom. We assume that the clocks at each location are started with the passage of a reference vehicle going from D to U. For this reason, the vehicle trajectories are skewed; i.e., vertical lines from D to U and downward slanting lines spanning 2T time units from U to D. Note how the vertical and slanted trajectories bisect each other. This ensures that all vehicles enter safely with a clearance of T =2.5 < l = 2.3. Immediately after an exit the next vehicle enters with the minimum clearance of s = 1.8.   Thus, the headway between successive entries at D is *T+s* = 2.5+1.8 = 4.3. Therefore, the unidirectional flow in this direction is 1/(4.3) = 0.23 [tu].  Consideration shows that this flow is at capacity because no more vehicles can be inserted in this diagram without violating the minimum headway or the safety rules. Now, since the system is symmetric, the same capacity arises in the opposite direction, as can also be seen from the figure. Therefore, the capacity point illustrated by the figure is (0.23, 0.23). Note how this matches the cluster point of Figure 3 for T=2.5.

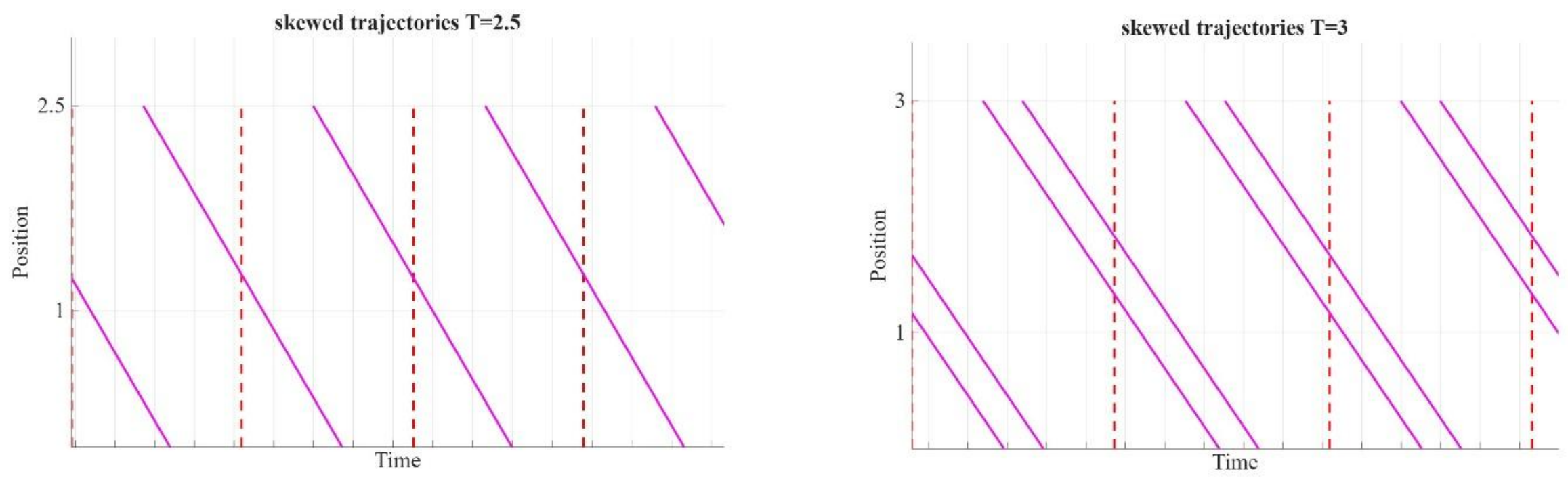

Figure 4: Skewed trajectories in mode A. Left pane: T= 2.5 (1 vehicle per platoon). Right pane: T = 3 (1 and 2 vehicles per platoon).

A figure similar to the one for T = 2.5 with bisecting trajectories and symmetric flows can also be drawn for medium blocks with T = 3, and the result is also a capacity outcome. This figure is not shown. With T=3, however, two other asymmetric solutions are also possible. The first of these is shown on the right pane of Figure 4. Note how two-car platoons flow from U to D and single cars from D to U.  The platoon trajectories have been slightly displaced horizontally, so the centerlines of the two-car platoons do not bisect the single car trajectories. Instead, the platoon centerlines cross the single vehicle trajectories slightly above or below the middle. Thanks to the platoons, the flow from U to D is higher than from D to U. The second asymmetric solution mirrors the first, with platoons flowing from D to U and single cars from U to D. Hence, there are three Mode A solutions in total: one with symmetric flows and two asymmetric. The reader can easily verify that the capacity points of these three solutions exactly match the three clusters observed below the capacity boundary in Figure 3 for T = 3.

For longer blocks there are more ways in which multiple vehicles can be simultaneously in the block, sometimes involving long platoons, which induce high capacities. These higher capacities can be seen in the simulations of Figure 3 for T=10 and T = 200.

### *3.3.3 Capacity of two bottlenecks under DP: Mode B*

The top pane of Figure 5 below shows a small-scale (low resolution) time-space diagram of this case. The differently shaded regions correspond to two phases with unidirectional flows in alternating directions so vehicle trajectories never cross. The solid shading denotes downward flow from U to D and the banded shading upward flow from D to U. The two sawtooth curves on the top and bottom of the time space diagram depict the length of the queues at U and D as function of time. Note how these queues grow when they are blocked by exiting traffic and dissipate when the exit flow ceases. Vehicles enter when they are unblocked by exiting vehicles. This occurs, both, when they arrive without a queue (passing through at a rate less than 1), and when their queue is discharging (at rate 1), which indeed block the blocking traffic from the other direction. The two alternating phases are separated by a phase transition, which is not shown in the top pane of the figure, but shall be described shortly.

During Phase I, bottleneck U discharges a diminishing queue, while D is blocked by these discharges so its queue grows. Flow goes downwards.

During Phase II the bottleneck roles are reversed: U is blocked and its queue grows, while D discharges a diminishing queue. Flow goes upwards.

Both phases are temporary because the diminishing queues eventually vanish. The result is that Phase I transitions into Phase II and the two phases continue to flip flop from then on.

The transition mechanism from II to I is depicted on the large-scale (high resolution) time-space diagram on the bottom pane of Figure 5, which was produced by a simulation.

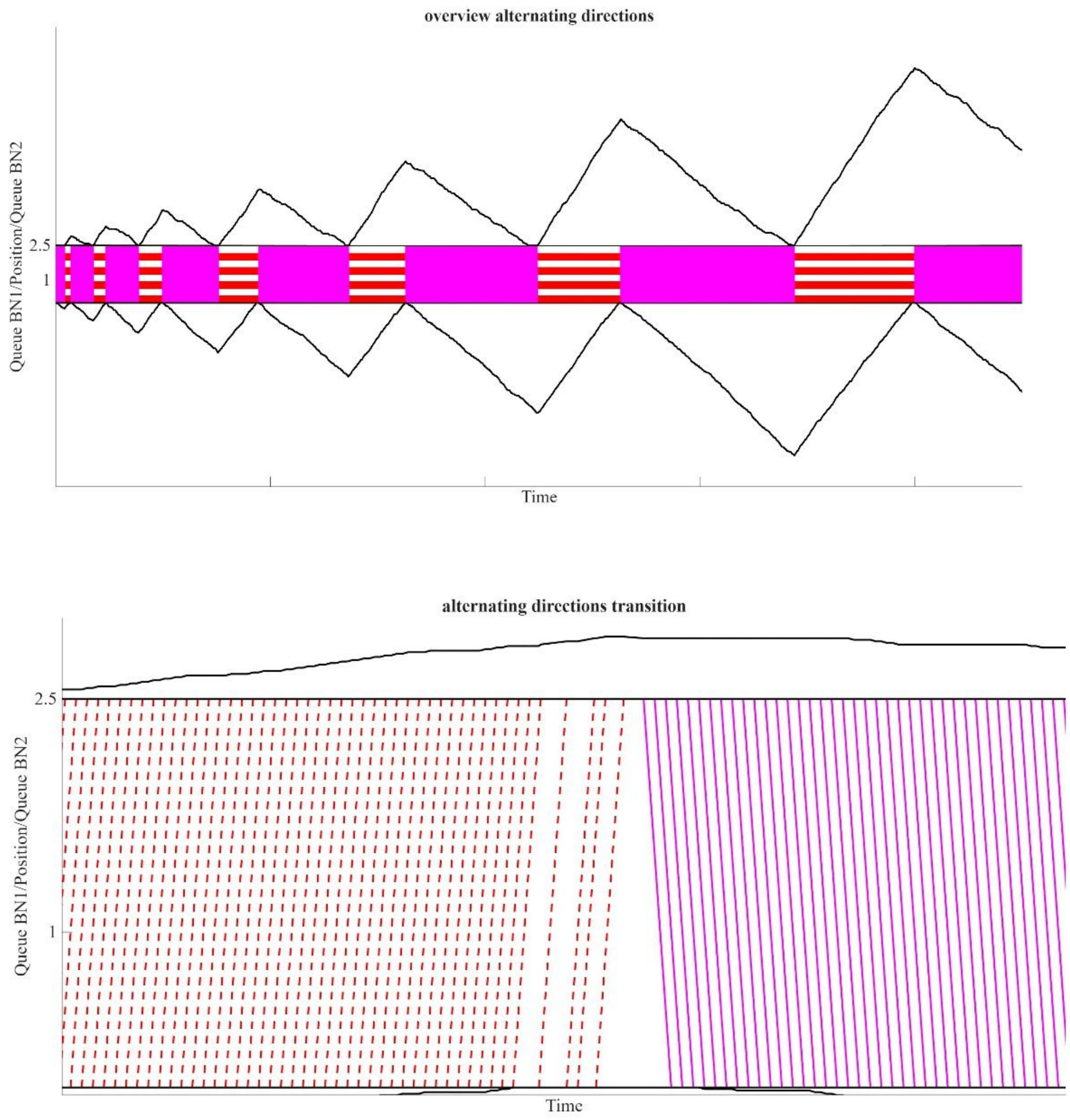

Figure 5: Illustration of Mode B. Top pane: Low resolution depiction of successive phases with alternating unidirectional flows and growing transient queues. Bottom pane: Large-scale depiction of the transition mechanism from Phase II to Phase I; vehicle trajectories are not skewed.

The transition begins when the queue at D dissipates. After this moment the arrivals at D pass through without delay. They do so as a series of platoons separated by gaps with an average flow, $a_D = d_D < 1$. At the other end, the arriving platoons force the queue to discharge intermittently at a rate $d_U < 1$. During this period of intermittence, the combined bidirectional flow is such that $d_U + d_D < 1$. This state of affairs continues until a "supercritical" gap with no traffic appears at D. The critical gap must be sufficiently long to allow the discharges it unblocks from U to arrive at D before the expiration of the gap, thereby blocking future arrivals at D. As shown by the figure, this blocking event marks the beginning of Phase I with downward flow only. (It is worth remembering that the critical gap to allow the transition must be longer than T.)

The same logic shows that Phase I must eventually end and transition back to Phase II.

This oscillation between phases I and II can go on forever. Because the block is oversaturated, the average transient queue during each phase grows without bound with each oscillation, as shown by the top pane of Figure 5. As these transient queues grow bigger and bigger, they take longer and longer to dissipate, so the phase durations also grow without bound.

Properties of Mode B

Note that the combined output flow from the two bottlenecks is $(d_U+d_D) = 1$ when in the system is either in Phase I or Phase II. During the phase transitions, the combined flow is $(d_U + d_D)<1$.

Also note that while the durations of Phases I and II increase without bound, the expected duration of a phase transition is fixed—as the bottom pane of Figure 5 shows, it is equal to the expected time to the appearance of a critical gap, plus $2T+ b$. This fixedness implies that the fraction of time that the system spends in Phases I and II (with $d_1 + d_2 = 1$) tends to 1 over time. This is as good an outcome as one could achieve with two perfectly controlled and coordinated traffic signals.

These ultra-high-capacity outcomes can be clearly seen in Fig. 3, for T=2.5 and T = 3. Similar outcomes do not appear in the panes for T=10 and T=200 because for large T, the critical gap needed to trigger the phase transitions is so long that it never pops up in our simulations.

## 3.4. The capacity with more than two bottlenecks

We reasonably propose that an arrival pair $(a_1, a_2)$ is undersaturated for a street with more than two bottlenecks if it is undersaturated for all its blocks. We claim that this is true for both FIFO and DP, as discussed below. A limited number of simulations confirm this conjecture.

FIFO: In this case the capacity of the street is expected to be the capacity of a single bottleneck as given by (3) and (4).

DP: In this case we recommend dividing the street near its middle in two parts, each including approximately the same number of bottlenecks, and then assigning priority at each bottleneck to the direction pointing away from the dividing point. With this assignment, there are no blocks with converging priorities. There is one block with priority to exiting traffic—the block straddling the dividing point--and the rest have "coincident" priorities. We reasonably expect the capacity outcomes of the street to be those of its middle block, as described in Sec. 3.3.3. As a practical matter, to make sure that people do not speed while traveling along the part of the street with coincident priorities, we recommend forcing the vehicles traveling these sections to zigzag sharply by placing the bottleneck barriers close to each other, making them wide, and alternating their positioning on the left and right sides of the street.

## 4. Conclusions and discussion

This paper analyzed the effect of one-vehicle-wide bottlenecks on a bidirectional traffic stream. We analyzed delays, capacity and traffic patterns for two different priority schemes: FIFO and directional priority (DP). We found that for low demand and any number of bottlenecks, FIFO causes less delay than DP.

For high demand, capacity is the main determinant of delay—the lower the capacity the higher the delay. Therefore, we paid quite a bit of attention to capacity. We found that for a single bottleneck, FIFO has greater capacity when the flows are unbalanced (see Fig. 2).

We also studied the capacity of two bottlenecks in series. For this case, the capacity of FIFO is the same as that of a single bottleneck. For the case of DP, this is not the case; curiously, for most instances simulated, the capacity turns out to be *higher* than for a single bottleneck.

The reason for the increased capacity with a second bottleneck is that platoons are spontaneously formed, reducing unused time at the bottleneck(s). Section 3.3 discussed several patterns which might occur. Some lead to a very efficient traffic flow with almost no unused time. The resulting capacity is comparable to that produced by two perfectly controlled and coordinated traffic signals.

This paper analyzed the theoretical properties, and took a first step in deriving different patterns that might arise and comparing different priority schemes. It used few measurements on operations. Future work should verify the operations in real life conditions and under more relaxed assumptions (driver heterogeneity, voluntary yielding, bounded acceleration).

***APPENDIX A*: Proof of Eq. (5) for the capacity formula for a DP bottleneck**

It follows from our assumptions that the mean headway preceding a platoon leader is $E(H_L) = 1+1/a_1$ and the mean headway preceding a follower is $H_F = 1$.

As a preliminary step, let us find the flow of platoon leaders $a_L$ as a function of the total, $a_1$. To do this, express the known mean of all the headways ($E(H) = 1/a_1$) as the weighted average of the mean headways for leaders and followers: $E(H_L) = 1+1/a_1$ and $E(H_F) = 1$.

Since the fraction of leaders is $a_L/a_1$ and that of followers is $(1-a_L/a_1)$, this weighted average is: $(a_L/a_1)(1+1/a_1)+(1-a_L/a_1) \equiv 1/a_1$. Now, solve for $a_L$ to find the flow of platoon leaders:

$$a_L = a_1(1-a_1). \qquad \text{(A1)}$$

Next, let us look for the expected number of slots available to serve secondary vehicles in each leader's headway. To do this, let K be a random variable denoting said number of slots, and $p_k$ the probability of the event $(K = k)$, where k = 0, 1, 2…

To fit a vehicle, a headway between platoons must last for at least b = l+s [tu] (s after the last vehicle of the first platoon and l before the arrival of the second platoon. Clearly, $K = 0$, if $H = 1+E < b$, so we have $p0 = \Pr(E < b-1) = 1-\exp(-a_1[b-1])$. Now, to simplify derivations, let $\exp(-a_1) \equiv \rho$ so we can write: $p0 = 1 - \rho^{[b-1]}$.

Let us now find $p_k$ for $k \geq 1$. To this end, first define the cumulative probability to have space for at least k vehicles, $P_k \equiv \Pr(K \geq k, \text{ for } k \geq 1)$. Note, this event occurs if and only if $([1+E] > [b+(k-1)])$; i.e., if and only if $(E > b+(k-2))$. Since E is negative exponential with mean $1/a_1$, we can write $P_k = \exp(-a_1[b+(k-2)]) = \rho^{[b-2]}\rho^k$, where $k \geq 1$. Now note that the probability to have space for exactly k vehicles, $p_k$, equals the probability that it fits at least k, but does not fit more than k: $p_k = P_k - P_{k+1}$, which simplifies to: $p_k = \rho^{[b-2]}\rho^k - \rho^{[b-2]}\rho^{k+1} = \rho^{[b-2]}(\rho^k - \rho^{k+1}) = \rho^{[b-2]}(1-\rho)\rho^k$.

Since the sequence $\{p_k\}$ is the probability mass function of K, we can write: $E(K)=kp_k = \sum_k k[\rho^{[b-2]}(1-\rho)\rho^k] = [\sum_{k\geq 1} k\rho^k][\rho^{[b-2]}(1-\rho)]$. The term for k = 0 has been omitted from the second summation because it is zero. The last expression can be simplified using the identity $[\sum_{k\geq 1} k\rho^k] \equiv \rho(1-\rho)^{-2}$, which is easily proven.[2] Thus, we can write: $E(K) = [\rho(1-\rho)^{-2}][\rho^{[b-2]}(1-\rho)] = \rho^{[b-1]}/(1-\rho)$. In terms of our original parameters, the expression becomes:

$$E(K) = \exp[-a_1(b-1)] / [1-\exp(-a_1)]. \qquad \text{(A2)}$$

Finally, note that the flow of available slots for a secondary vehicle in the priority stream is $a_L E(K)$ and that the maximum feasible secondary flow $c(a_1)$ is achieved when every such available slot is filled by a low priority vehicle; i.e., when the secondary flow is $c(a_1) = a_L E(K)$. Now insert (A1) and (A2) in the RHS and note the result is Eq. (5) of the text. QED

---

[2] *Proof*: We know that $[\sum_{k\geq 1} \rho^k] = \rho/(1-\rho)$. (This is the geometric series identity.) Now, take derivatives on both sides of the equality with respect to $\rho$ and obtain: $[\sum_{k\geq 1} k\rho^{k-1}] = 1/(1-\rho)^2$.The proof is complete by multiplying both sides of this last equality by $\rho$, which yields: $[\sum_{k\geq 1} k\rho^k] = \rho/(1-\rho)^2$, as claimed.